\documentclass[showpacs,preprintnumbers,amsmath,amssymb,prd]{revtex4}

\usepackage{graphicx}
\usepackage{dcolumn}
\usepackage{bm}

\newcommand{\mpc}{\, {\rm Mpc}}

\newcommand{\hmpc}{\, h^{-1} \mpc}
\newcommand{\ihmpc}{\, h\, {\rm Mpc}^{-1}}

\newcommand{\lyaf}{Ly$\alpha$ forest}

\newcommand{\vq}{\mathbf{q}}
\newcommand{\vx}{\mathbf{x}}
\newcommand{\vk}{\mathbf{k}}
\newcommand{\tbtwo}{\tilde{b}_2}

\newcommand{\orderfour}{\mathcal{O}\left(\delta_1^4\right)}

\begin{document}

\title{Clustering of dark matter tracers:  renormalizing the bias parameters}

\author{Patrick McDonald}
\email{pmcdonal@cita.utoronto.ca}
\affiliation{Canadian Institute for Theoretical Astrophysics, University of
Toronto, Toronto, ON M5S 3H8, Canada}

\date{\today}

\begin{abstract}

A commonly used perturbative method for computing large-scale clustering of
tracers of mass density, like galaxies, is to model the tracer density field as
a Taylor series in the local smoothed mass density fluctuations, possibly 
adding a stochastic component.  I suggest a set of parameter redefinitions, 
eliminating problematic perturbative correction terms, that should represent a 
modest improvement, at least, to this method.  As presented here, my method can
be used to compute the power spectrum and bispectrum to 4th order in initial 
density perturbations, and higher order extensions should be straightforward.  
While the model is technically unchanged at this order, just reparameterized, 
the renormalized model is more elegant, and should have better convergence 
behavior, for three reasons:  First, in the usual approach the effects of 
beyond-linear-order bias parameters can be seen at asymptotically large scales,
while after renormalization the linear model is preserved in the large-scale 
limit, i.e., the effects of higher order bias parameters are restricted to 
relatively high $k$.  Second, while the standard approach includes smoothing to 
suppress large perturbative correction terms, resulting in dependence on the 
arbitrary cutoff scale, no cutoff-sensitive terms appear explicitly after my 
redefinitions (and, relatedly, my correction terms are less sensitive to 
high-$k$, non-linear, power).  Third, the 3rd order bias parameter disappears 
entirely, so my model has one fewer free parameter than usual (this parameter 
was redundant at the order considered).  This model predicts a small
modification of the baryonic acoustic oscillation (BAO) signal, in real space,
supporting the robustness of BAO as a probe of dark energy, and providing a 
complete perturbative description over the relevant range of scales.
 
\end{abstract}


\maketitle

\section{Introduction}

In the past year, significant progress has been made in using perturbation 
theory (PT) to calculate the large-scale clustering of collisionless mass in 
the Universe, to
the point where essentially perfect calculations of the quasi-linear clustering
may soon be available \cite{2006PhRvD..73f3520C,2006PhRvD..73f3519C,
2006astro.ph..6028M}.  
This motivates a fresh look at the bias models that are needed to couple these
calculations to most observable tracers of large-scale mass density, e.g., 
galaxies \citep{2004ApJ...606..702T,2006MNRAS.366..189S}, the \lyaf\
\citep{2005ApJ...635..761M,2006ApJS..163...80M,2006MNRAS.365..231V},
galaxy cluster/Sunyaev-Zel'dovich effect (SZ) measurements
\citep{2005astro.ph.11060D}, and possibly
future 21cm surveys \citep{2005MNRAS.364..743N}.  Baryonic acoustic 
oscillation surveys aimed at probing dark energy
\citep{1998ApJ...504L..57E,
1998ApJ...496..605E,2001ApJ...557L...7C,2003astro.ph..1623E,
2003ApJ...594..665B,2003PhRvD..68h3504L,2003ApJ...598..720S,
2004ApJ...615..573M,2005ApJ...631....1G,2005astro.ph..7457G,
2005MNRAS.357..429A,2005MNRAS.363.1329B,2006MNRAS.365..255B,
2006MNRAS.366..884D,2006astro.ph..7122M,2006astro.ph..4075J}, in particular, 
fall in the range of 
scales where perturbation theory may be most useful.

Probably the most straightforward, commonly used, model for the bias when 
coupled to perturbation theory is to write the tracer density, $\delta_g$, as 
a Taylor series in the mass density fluctuations, $\delta$ 
\citep{1993ApJ...413..447F,1998MNRAS.301..797H}, i.e.,
$\delta_g = \sum_{i} b_i \delta^i/i!$.
Note that I will often refer to the tracer as galaxies (this explains the
subscript $g$), but essentially everything in the paper could be applied to 
any tracer.
The terms in this series generally will not decrease in size, so $\delta$ is 
usually taken to be a smoothed version of the density field, to reduce the size
of its fluctuations.  When 
$\delta$ is computed using perturbation theory for gravitational clustering
\citep{1980lssu.book.....P,1981MNRAS.197..931J,1983MNRAS.203..345V,
1984ApJ...279..499F,1986ApJ...311....6G,1994ApJ...431..495J,
1996ApJ...473..620S,2002PhR...367....1B}, clustering of the tracers is in 
principle fully described.
\citep{1998MNRAS.301..797H} showed that applying this approach to compute the
power spectrum up to 4th order in the initial density perturbations (the lowest 
order that gives a correction to linear theory) leads to interesting, but not
completely satisfactory results.  They found that the 2nd order bias term 
(i.e., $b_2$) produces a white ($k$-independent) contribution to the power on 
large scales.  Furthermore, both the 2nd and 3rd order bias parameters 
contribute
terms to the effective large-scale bias, defined as the ratio of the galaxy
to mass power.  These terms all depend on the scale
of the smoothing applied to the density field.  Note that the most 
straightforward expectation is that the smoothing scale should be quite large,
as one can see by considering the measured value of rms density fluctuations 
in $8\hmpc$ radius spheres, $\sigma_8\simeq0.85$ 
\citep{2006JCAP...10..014S}.  This
level of smoothing dramatically affects the power on relatively large 
scales, e.g., an $8\hmpc$ radius top-hat smoothing suppresses the power at 
$k=0.2\ihmpc$ by almost 50\%.

It may be 
intrinsically interesting, when studying galaxy formation, to determine the 
coefficients $b_i$ in the series $\delta_g = \sum_{i} b_i \delta^i/i!$,
and to see generally how well 
this model works, as a function of smoothing scale.
However, I will suggest a cleaner repackaging of the model, for the 
specific purpose of large-scale structure phenomenology, which has the 
desirable properties of 
preserving the form of the linear theory model at very large scales (i.e., 
the power at small $k$ is described by a single linear theory bias parameter 
and a single shot-noise parameter), and being free of any explicit 
smoothing scales. 
The approach I take is more or less precisely the same as the one used in  
classic renormalization of quantum field theories.  When perturbation theory is
pursued naively, infinities appear, but only in ways that can be reinterpreted
as corrections to the values of parameters of the model (e.g., particle 
masses, analogous to our linear model parameters).  The infinities are removed 
by simply redefining the free parameter in the model to be the most directly 
observable quantity (e.g., the mass you would measure by weighing the particle,
or the ratio of large-scale galaxy to mass power), which would be given 
by a series of terms if one continued to use the original parameterization 
\citep{Peskin:1995ev}.  

Section \ref{seccalc} of this paper gives my primary calculations, followed by 
some discussion in section \ref{secdiscuss}.  
Before proceeding I reiterate the main goals:  (1) To formulate the bias model 
in a way that does not require
arbitrary smoothing scales, while simultaneously being insensitive
to the small-scale, highly non-linear regime.  (2) To formulate the model in
a way that preserves the linear theory model on very large scales (low-$k$),
i.e., higher order bias terms only affect the relatively high-$k$ power. 
To be honest, I should also note here that, while I refer throughout the paper
to the approach of
\citep{1998MNRAS.301..797H} as the ``usual'' or ``standard'' approach, this
perturbative approach to the galaxy power spectrum (including beyond-linear
corrections) has not to my knowledge 
actually been used to interpret real data.  However, between 
improvements in perturbation theory and the need to 
interpret increasingly precise observations, the time for this kind of approach
may have arrived \citep{2006astro.ph..4075J}.

\section{Calculation \label{seccalc}}

We are interested in the statistics of the real-space galaxy density field,
$\delta_g(\vx)=\rho_g(\vx)/\bar{\rho}_g-1$, with Fourier transform
$\delta_g(\vk)=\int d^3\vx~\exp(i \vk\cdot \vx)~\delta_g(\vx)$,
where $\vx$ is the comoving position.
I will compute the power spectrum, defined by 
$\left<\delta_g(\vk)\delta_g(\vk^\prime)\right>=
(2\pi)^3\delta^D(\vk+\vk^\prime)P_g(k)$,
and bispectrum, 
$\left<\delta_g(\vk_1)\delta_g(\vk_2)\delta_g(\vk_3)\right>=
(2\pi)^3\delta^D(\vk_1+\vk_2+\vk_3)B_g(k_1,k_2,k_3)$.  
I am going to use perturbation theory to describe the mass density 
fluctuations, $\delta=\delta_1+\delta_2+\delta_3+\orderfour$, where 
$\delta_n$ is of order $\delta_1^n$ \citep{1980lssu.book.....P,
1981MNRAS.197..931J,1983MNRAS.203..345V,1984ApJ...279..499F,
1986ApJ...311....6G,1994ApJ...431..495J,1996ApJ...473..620S}.
I start by writing the galaxy density as a Taylor series in 
$\delta$ \citep{1993ApJ...413..447F,1998MNRAS.301..797H}:
\begin{equation}
\rho_g(\delta)=\rho_0+\rho^\prime_0~\delta+
\frac{1}{2}\rho_0^{\prime\prime}~\delta^2+
\frac{1}{6}\rho_0^{\prime\prime\prime}~\delta^3+\epsilon+\orderfour~,
\label{eqtaylor}
\end{equation}
stopping at 3rd order because this what is needed to compute the lowest order
bispectrum
and the first non-linear correction to the power spectrum.  I have added an
uncorrelated noise variable $\epsilon$ to represent shot-noise and other 
randomness in
the galaxy-mass relation that appears as white noise on large scales
\cite{1999ApJ...520...24D}.
(Generally, anything that affects the correlation function only at small
separations will appear as a change in white noise level in the low-$k$ power 
spectrum.  It is important to model this noise, rather than trying to avoid it 
by working with the correlation function restricted to large separations,
because it contributes to the measurement errors on either statistic.)
The variance of $\epsilon$ is $\left<\epsilon^2\right>=N_0$. 
The Taylor series coefficients and noise amplitude are effectively free 
parameters as long as we do not have a fully predictive galaxy formation 
model.  Usually $\delta$ is understood to be a smoothed version of the 
density field, in order to force it to be small enough to justify the Taylor
expansion \citep{1993ApJ...413..447F,1998MNRAS.301..797H}.  I do not need to 
introduce this smoothing explicitly. 

As a warm-up calculation, introducing the basic idea of renormalization that I 
will use later, I compute the mean density of galaxies:
\begin{equation}
\left<\rho_g\right>=\rho_0+
\frac{1}{2}\rho_0^{\prime\prime}\left<\delta^2\right>+ \orderfour
\end{equation}
Note that $\left<\delta^3\right>$ is not zero, but it is $\orderfour$ 
($\left<\delta\right>=0$ by definition).
In absence of a cutoff (smoothing scale), 
$\left<\delta^2\right>\equiv\sigma^2$ is infinite, or at least potentially 
large.  Equivalently, if one introduces an arbitrary cutoff,
the value of $\sigma^2$ will be sensitive to the cutoff.
This is not a problem -- I simply 
define the observed mean galaxy density to be $\bar{\rho}_g= \rho_0+
\frac{1}{2}\rho_0^{\prime\prime}\sigma^2$.  I can then
remove $\rho_0$ from Eq. (\ref{eqtaylor}) in favor of $\bar{\rho}_g$, 
obtaining 
\begin{equation}
\rho_g(\delta)=\bar{\rho}_g+\rho^\prime_0~\delta+
\frac{1}{2}\rho_0^{\prime\prime}~\left(\delta^2-\sigma^2\right)+
\frac{1}{6}\rho_0^{\prime\prime\prime}~\delta^3+\epsilon+\orderfour~.
\label{meanrenormrhog}
\end{equation}
This calculation has not been very profound
(it is so obvious that it is often done with little or no comment 
\citep{1993ApJ...413..447F,1998MNRAS.301..797H}), but it does provide a very 
simple example of the idea of absorbing certain kinds of bad behavior in a 
perturbative expansion into a redefinition of the parameters, in particular
replacing a parameter in the original model by a more directly observable 
quantity that would otherwise be calculated as a function of the original 
parameters.  
Sufficiently aggressive smoothing could of course render the 2nd term in the
calculation of the mean density smaller than the
first, but this is partly missing the point:  there is no reason to allow the
mean galaxy density to acquire corrections at each order when a simple 
redefinition of the original, relatively meaningless, Taylor series 
coefficients 
can guarantee that this fundamental observable is always directly represented 
by a single parameter in the model, with no need for smoothing.

Note that when I say above that $\sigma^2$ is ``infinite or potentially 
large'', the
uncertainty in this statement comes from the form of perturbation theory used 
to compute the mass power spectrum, e.g., standard PT
\citep{1980lssu.book.....P,1981MNRAS.197..931J,1983MNRAS.203..345V,
1984ApJ...279..499F,1986ApJ...311....6G,1994ApJ...431..495J,
1996ApJ...473..620S,2002PhR...367....1B}, the renormalized PT of 
\cite{2006PhRvD..73f3520C,2006PhRvD..73f3519C}, or the renormalization 
group (hereafter RG)-improved PT of \cite{2006astro.ph..6028M}, and whether 
and how much smoothing is 
applied.  I will discuss this more below, but, in any case, it is clear that 
$\sigma^2$ is sensitive to the details of one's treatment of the non-linear 
regime, so it is desirable to eliminate it from the final results of the 
calculation if possible.

I now move on to fluctuations around the mean galaxy density, defining
\begin{equation}
\delta_g\left(\delta\right)=\frac{\rho_g\left(\delta\right)-\bar{\rho}_g}
{\bar{\rho}_g}=c_1 \delta + \frac{1}{2}c_2 \left(\delta^2-\sigma^2\right)+
\frac{1}{6}c_3 \delta^3+\epsilon+\orderfour 
\end{equation}
($\epsilon$ has been rescaled).
The galaxy correlation function is
\begin{equation}
\xi_g\left(\left|\vx_a-\vx_b\right|\right)=
\left<\delta_g(\vx_a) \delta_g(\vx_b)\right>=
c_1^2\left<\delta_a \delta_b\right>+
\frac{1}{3}c_1 c_3 \left<\delta_a \delta_b^3\right>+
\frac{1}{4}c_2^2\left(\left<\delta_a^2 \delta_b^2\right>-\sigma^4\right)+
c_1 c_2 \left<\delta_a \delta_b^2\right>+\left<\epsilon_a\epsilon_b\right>+..., 
\end{equation}
where $\delta_a \equiv\delta(\vx_a)$.
Using the fact that the 4th order terms can be treated as Gaussian at the 
order I am considering, and the linear or non-linear correlation function
can be used interchangeably in these terms,   
\begin{equation}
\xi^{ab}_g=
c_1^2\xi_{ab}+
c_1 c_3 \sigma^2 \xi_{ab}+
\frac{1}{2}c_2^2\xi^2_{ab}+
c_1 c_2 \left<\delta_a \delta_b^2\right>+N_0\delta^D_{ab}+...~.
\end{equation}
We see that a term has arisen, $c_1 c_3 \sigma^2 \xi_{ab}$, that has the
form of a correction to the linear theory bias, in the sense that there is a 
$k$-independent factor multiplying the correlation function.  It is 
divergent, or at least
potentially large, but this term can be eliminated by redefining the bias to 
include it,
similar to what I did with the mean density.  I will not redefine the bias yet,
however, because another similar term will arise from 
$\left<\delta_a \delta_b^2\right>$, which I have not evaluated yet because it
requires beyond-linear perturbation theory for the density fluctuations.

To go farther, it is simplest to move to Fourier space, where the galaxy 
density field is
\begin{equation}
\delta_g(\vk)=c_1 \delta_\vk + 
\frac{1}{2}c_2\int \frac{d^3 \vq}{\left(2 \pi\right)^3} 
\delta_\vq \delta_{\vk-\vq} +
\frac{1}{6}c_3\int \frac{d^3 \vq_1}{\left(2 \pi\right)^3}
\frac{d^3 \vq_2}{\left(2 \pi\right)^3}
\delta_{\vq_1}\delta_{\vq_2} \delta_{\vk-\vq_1-\vq_2} +
\epsilon_\vk+\orderfour ~.
\label{eqdeltagFourier}
\end{equation}
The mass density field, up to 3rd order, is  
\begin{equation}
\delta_\vk = \delta_1(\vk)+
\int \frac{d^3 \vq}{\left(2 \pi\right)^3}\delta_1(\vq) \delta_1(\vk-\vq)
J_S^{(2)}(\vq,\vk-\vq)+
\int \frac{d^3 \vq_1}{\left(2 \pi\right)^3}
\frac{d^3 \vq_2}{\left(2 \pi\right)^3}
\delta_1(\vq_1) \delta_1(\vq_2) \delta_1(\vk-\vq_1-\vq_2)
J_S^{(3)}(\vq_1,\vq_2,\vk-\vq_1-\vq_2)+...
\label{eqdelta3rdorder}
\end{equation}
where
\begin{equation}
J_S^{(2)}(\vk_1,\vk_2)=
\frac{5}{7}+\frac{1}{2}\frac{\vk_1\cdot\vk_2}{k_1 k_2}
\left(\frac{k_1}{k_2}+\frac{k_2}{k_1}\right)+\frac{2}{7}
\left(\frac{\vk_1\cdot\vk_2}{k_1 k_2}\right)^2~, 
\end{equation}
and see Eq. 11 of \cite{1998MNRAS.301..797H} for $J^{(3)}$ (it will not 
actually be used in this paper).
Combining Eq. (\ref{eqdeltagFourier}) and Eq. (\ref{eqdelta3rdorder}), we find 
the power spectrum 
\begin{eqnarray}
P_g(k)&=&
N_0+[c_1^2+ 
c_1 c_3 \sigma^2 +\frac{68}{21}c_1 c_2 \sigma^2] P(k)+\nonumber \\
& &\frac{1}{2}c_2^2 \int \frac{d^3\vq}{(2 \pi)^3} P(q) 
P(\left|\vk-\vq\right|)+ \nonumber \\
& &2~c_1 c_2 \int \frac{d^3\vq}{(2 \pi)^3} 
P(q) P(\left|\vk-\vq\right|)
J^{(2)}_S(\vq,\vk-\vq)+...
\label{eqbarepower}
\end{eqnarray}
Note that, if the $\delta$'s in the original Taylor series were smoothed,
the term $\frac{68}{21}c_1 c_2 \sigma^2$ would not take the form of a simple
$k$-independent bias, except in the limit that the smoothing scale is very 
small (see \citep{2006astro.ph..9547S} for a different approach in which 
extreme smoothing is applied).

Eq. (\ref{eqbarepower}) suggests the following redefinition of the linear bias
parameter:
\begin{equation}
b_1^2=c_1^2+c_1 c_3 \sigma^2 + \frac{68}{21} c_1 c_2 \sigma^2~.
\end{equation}
\cite{1998MNRAS.301..797H} identified $b_1^2$ as defined here as an effective
bias.  The difference in approach is that I am suggesting that $b_1$ should 
now be treated
as the free parameter of the model, with $c_1$ and $c_3$ eliminated from 
Eq. (\ref{eqbarepower}) by substitution (I write the result in Eq. 
\ref{eqfinalpower} below, after some further redefinitions to be discussed 
next).
Note that, because the last term in Eq. (\ref{eqbarepower}) is already 
$\orderfour$, $c_1$ in that term can be freely replaced by $b_1$, which is 
equal to $c_1$ at lowest order.
Note also that $b_1$ includes the only appearance of $c_3$, so the third order 
bias has disappeared as an independent parameter.

Eq. (\ref{eqbarepower}) contains a final divergence, or at least 
high-$k$-sensitive term,  proportional to the following:
\begin{equation}
\int \frac{d^3\vq}{(2 \pi)^3} P(q)
P(\left|\vk-\vq\right|)~.
\label{eqdivergentnoiseterm}
\end{equation}
This integral diverges at high $q$ 
if the asymptotic logarithmic slope of the power spectrum is
$n_{\rm eff}(q\rightarrow \infty)\geq-1.5$.  
This is not the case for the standard linear theory calculation of the 
$\Lambda$CDM power spectrum; however, the fixed-point asymptote of the 
RG-corrected power spectrum of 
\cite{2006astro.ph..6028M} 
is $P(q)\propto q^{-1.4}$, which would give an infinite integral.  As argued 
by \cite{2006PhRvD..73f3520C,2006PhRvD..73f3519C,2006astro.ph..6028M}, the 
standard perturbation theory calculation based on using the usual linear theory 
result as the source for higher order terms, at all times and $k$, does not 
make much sense (and does not work very well when compared to numerical 
simulations), because at high $k$ the initial power spectrum is 
quickly completely erased.  Especially at late times, $n_{\rm eff}\simeq-1.4$ 
is much
more relevant.  While this is not a completely essential component of this 
paper,
the beyond-linear terms in Eq. (\ref{eqbarepower}) would most naturally be 
evaluated using the renormalized mass power spectrum of 
\cite{2006astro.ph..6028M} (in fact, in the approach of 
\cite{2006astro.ph..6028M} there is really no other option).
In the alternative approach of \cite{2006PhRvD..73f3520C,2006PhRvD..73f3519C} 
these terms
would probably be evaluated using the linear theory power as suppressed at high
$k$ by the renormalization of the propagator. 

In any case, at best the result for the term in Eq. 
(\ref{eqdivergentnoiseterm}) is sensitive to the treatment of the non-linear 
regime, which we would like to avoid.  Another problem with this term is that 
it is non-zero as 
$k\rightarrow 0$, i.e., it is a correction to the power spectrum at 
asymptotically large scales.  As discussed by \cite{1998MNRAS.301..797H}, this 
correction is 
$k$-independent at low $k$, so it looks like a correction to the shot-noise.
A basic philosophy of this paper is that we would like to preserve the simplest
linear theory model for galaxy clustering on large scales, rather than allowing
it to be corrected at each order in perturbation theory.
Therefore, we now perform our final renormalization, absorbing this potentially
divergent quantity,
evaluated at $k=0$, into the shot-noise term, which becomes
\begin{equation}
N=N_0+\frac{1}{2}c_2^2 \int \frac{d^3\vq}{(2 \pi)^3} P^2(q)~.
\end{equation}
After this piece is subtracted, the remaining integral, written as a Taylor
series around $k=0$,
\begin{equation}
\int \frac{d^3\vq}{(2 \pi)^3} P(q)
\left(P(\left|\vk-\vq\right|)-P(q)\right)\simeq 
\int \frac{d^3\vq}{(2 \pi)^3} P(q)
\left(-\mu k \frac{dP(q)}{dq}\right)+{\mathcal O}(k^2)~,
\end{equation}
(where $\mu = \vk\cdot \vq/k q$)
is clearly convergent for any reasonable power spectrum (in fact, the 
${\mathcal O}(k)$ term is zero after
the $\mu$ integration, making it even more convergent).

The 2nd order bias is not renormalized at this order, but for notational
compactness I define
\begin{equation}
\tilde{b}_2=\frac{c_2}{b_1}~.
\end{equation}
The final result for the power spectrum is:
\begin{equation}
P_g\left(k\right)=
b_1^2 \left[ P\left(k\right)+
\frac{\tilde{b}_2^2}{2}\int \frac{d^3\vq}{\left(2 \pi\right)^3} 
P\left(q\right)
\left[P\left(\left|\vk-\vq\right|\right)-P\left(q\right)\right]+
2~\tilde{b}_2 \int \frac{d^3\vq}{\left(2 \pi\right)^3} P\left(q\right)
P\left(\left|\vk-\vq\right|\right)
J^{\left(2\right)}_S\left(\vq,\vk-\vq\right) \right]+N~.
\label{eqfinalpower}
\end{equation}
We see that deviations from the traditional large-scale galaxy clustering 
model, in which the mass power is multiplied 
by a constant and white noise is added, are controlled by the single parameter,
$\tilde{b}_2$.
The first term (linear in $P(k)$) in Eq. (\ref{eqfinalpower}) must be 
evaluated using the non-linear mass power spectrum.
In standard perturbation theory, the $P$'s in the extra bias terms
(quadratic in $P(k)$) should be
understood to be the linear power, while in the RG approach
of \cite{2006astro.ph..6028M} they would be the renormalized power. 
In the approach of \cite{2006PhRvD..73f3520C,2006PhRvD..73f3519C} these $P$'s
would presumably be the initial power as evolved using the renormalized 
propagator.  Note that in some technical sense there is no difference between 
these options, at the order of calculation in this paper. 

\begin{figure}
\resizebox{\textwidth}{!}{\includegraphics{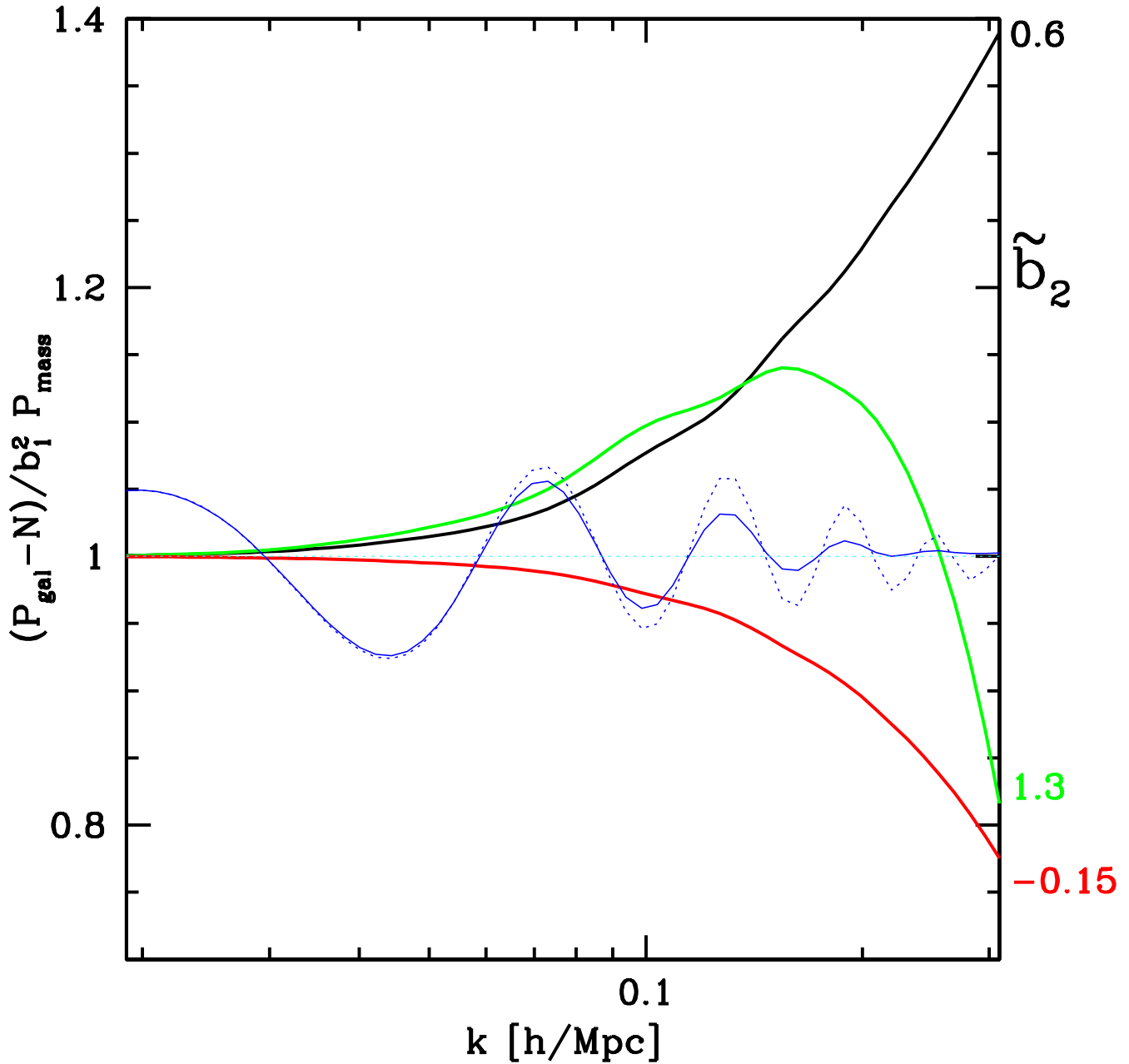}}
\caption{
Thick lines show $(P_g(k)-N)/ b_1^2 P_{\rm mass}(k)$ for $\tbtwo=0.6$ (black),
$\tbtwo=1.3$ (green), and $\tbtwo=-0.15$ (red).  For comparison, the thinner,
blue, wiggly lines show the ratio of mass power in a realistic model to the
power in a similar model with no baryonic acoustic oscillations (solid is 
non-linear power, dotted is linear).  
}
\label{figbasic}
\end{figure}

Fig. \ref{figbasic} shows the dependence of the power on $\tilde{b}_2$.
The cosmological model for this figure, plotted at $z=0$, was flat with 
$\sigma_8=0.85$, 
$n=0.96$, 
$\Omega_m=0.281$, 
$h=0.71$, and $\Omega_b=0.0462$ \citep{2006JCAP...10..014S}, with CMBfast 
\citep{1996ApJ...469..437S} used to generate the transfer function.
We see the desired convergence to the simple linear bias plus white noise model
on large scales.  The term $\propto \tbtwo^2$ is generally negative, while the
term $\propto \tbtwo$ has the sign of $\tbtwo$.  This means that when 
increasing
$\tbtwo$ from zero we obtain an increase in power at first, up to 
$\tbtwo\simeq 0.6$,
but then a decrease, especially at higher $k$, as the $\tbtwo^2$ term takes 
over.  While this bias term has a maximum, note that the simple ratio of  
galaxy to mass power does not, because extra white noise power can always be
added.
For increasingly negative $\tbtwo$ both terms are negative so the power 
decreases quickly (again, noise power can be added, so negative $\tbtwo$ does
not automatically mean that the bias as traditionally defined decreases).  
It is interesting to note that there are only hints of
features related to baryonic acoustic oscillations (BAO) in the ratio 
$(P_g(k)-N)/ b_1^2 P_{\rm mass}(k)$, at the 0.3\% 
level for the $\tbtwo=0.6$ case.  Combining their small size with the fact
that much of the effect appears to be a modification of the amplitude of the 
wiggles, rather than their position, suggests that 1\% distance measurements 
using BAO should not be significantly corrupted.  
Of course, these effects, the effect of broad-band 
$k$-dependence of the bias, and the modification already present in the 
non-linear mass power spectrum, should not be ignored when fitting 
observations.

\begin{figure}
\resizebox{\textwidth}{!}{\includegraphics{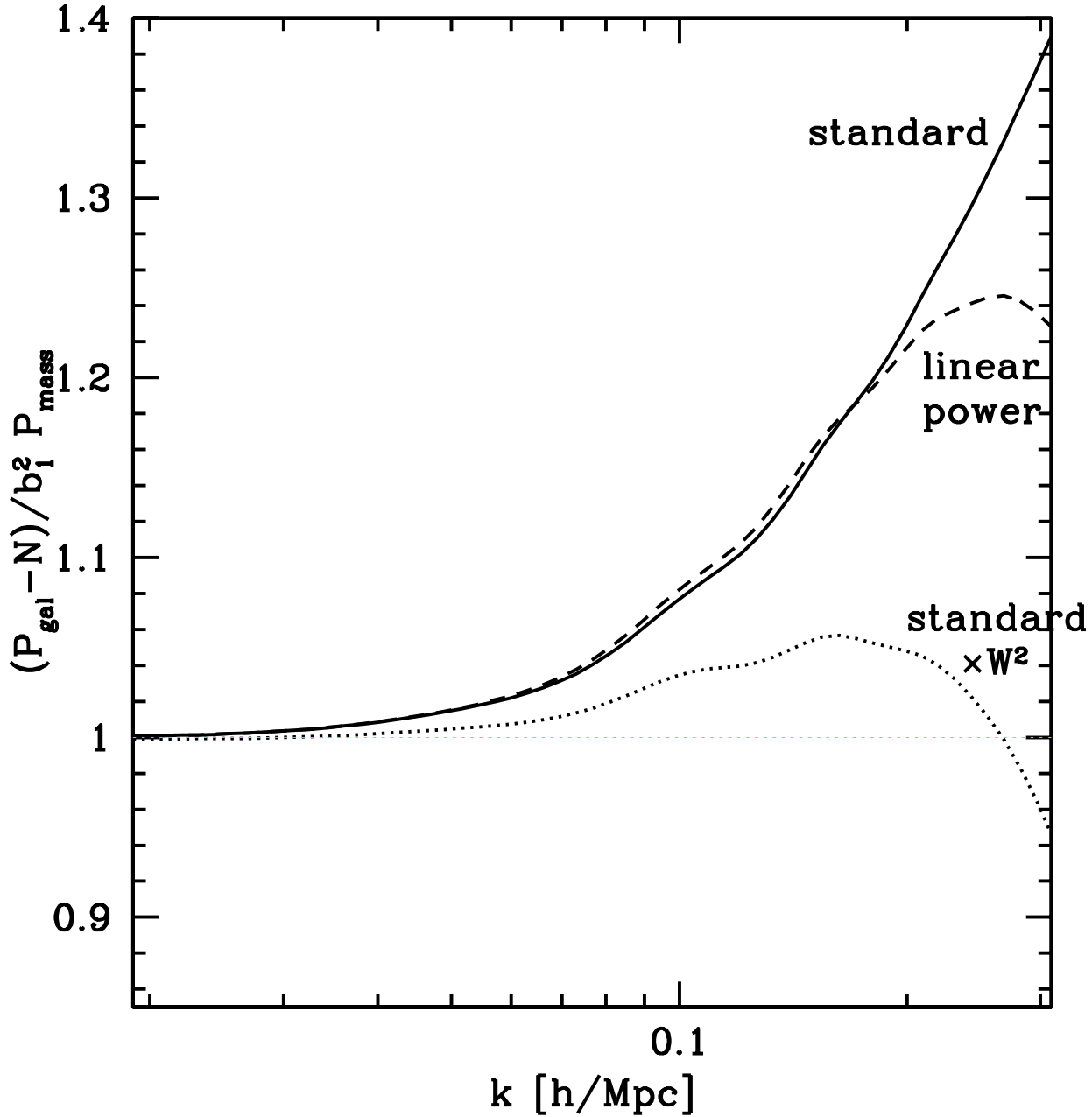}}
\caption{
$(P_g(k)-N)/ b_1^2 P_{\rm mass}(k)$ for $\tbtwo=0.6$.  The
solid curve is my standard calculation, where the RG-corrected (non-linear)
mass power spectrum has been
used to compute the bias terms.  To show the lack of sensitivity to
small scales, for the dashed curve the usual linear mass
power, which becomes dramatically different from the RG power with increasing
$k$, was used to compute the bias terms.  The dotted curve shows
the effect of a $2 \hmpc$ rms Gaussian smoothing applied to the
result of the standard calculation.
}
\label{fighighk}
\end{figure}

In Fig. \ref{fighighk} I explore the sensitivity of the higher order 
bias terms to high-$k$ power.    
For the standard results I have used the RG-improved power
spectrum of \cite{2006astro.ph..6028M} to compute these terms, but in this
figure
I compare to the result when I use the standard linear power spectrum.
We see that there is not much 
difference between the results for these cases, even though the RG
power is 37\% larger at $k=0.3\ihmpc$, more than a factor of 3 larger at 
$k=1\ihmpc$, and the asymptotic slopes at high $k$ are $-1.4$ vs. $\sim -3$.
This is a testament to how successful the inoculation against high-$k$ power
has been (recall that the unrenormalized bias-related corrections in the RG 
case would be 
literally infinite, and some of the corrections using standard PT nearly so).
I used 
$\tbtwo=0.6$ for the figure, but the results for larger or smaller $\tbtwo$
are similar.  An increase in sensitivity to small scale power with 
increasing $\tbtwo$ indicates that the term quadratic in $\tbtwo$ is more 
sensitive than the linear term. 
For reference, Fig. \ref{fighighk} also shows 
the effect of Gaussian smoothing with rms width 
$2 \hmpc$, which gives $\sigma=1.26$ for the model in the figures, i.e., 
not really perturbative.  
The direct power suppression for this smoothing
is already substantial, reaching 30\% at $k=0.3\ihmpc$.  

\subsection{Three-point function}

At the order in $\delta_1$ considered here, the three-point function 
(and presumably other
similar statistics that require only computation to this order)
\citep{1994ApJ...425..392F,1998ApJ...496..586S,2001PhRvL..86.1434F,
2002MNRAS.335..432V,
2003MNRAS.344..776M,
2005MNRAS.362.1363P,2005PASJ...57..709H,
2005MNRAS.361..824G,
2005MNRAS.364..620G,
2006astro.ph..5748R}
is described neatly by 
the renormalized model, with no non-trivial new effects.
In real space it is
\begin{eqnarray} 
\left<\delta_g(\vx_a)\delta_g(\vx_b)\delta_g(\vx_c)\right>=\zeta_g^{abc}&=&
c_1^3 \left<\delta_a \delta_b\delta_c\right>+\frac{1}{2}c_1^2 c_2
\left<\delta_a\delta_b\left(\delta_c^2-\sigma^2\right)+{\rm cyclic~
permutations}\right>+...\nonumber \\
&=& c_1^3 \zeta_{abc}+\frac{1}{2}c_1^2 c_2
\left(\xi_{ac}\xi_{bc}+{\rm cyclic}\right)+...
\end{eqnarray}
Since this is already $\orderfour$ we can freely substitute 
$c_1\rightarrow b_1$ to produce
\begin{equation} 
\zeta^{abc}_g=
b_1^3 \left[\zeta_{abc}+\frac{1}{2}\tilde{b}_2
\left(\xi_{ac}\xi_{bc}+{\rm cyclic}\right)\right]+...
\end{equation}
Deviations from the simplest model are again controlled entirely by 
$\tilde{b}_2$.

\section{Discussion\label{secdiscuss}}

The standard method for 
including biasing of tracers of mass density in 
perturbation theory calculations of large-scale clustering 
\citep{1993ApJ...413..447F,1998MNRAS.301..797H} uses five 
parameters to model the power spectrum and bispectrum up to 4th 
order in density perturbations, and requires at least one arbitrary cutoff 
parameter (smoothing scale).  
In the standard approach the effects of the beyond-linear order bias
parameters and the cutoff propagate to asymptotically large scales.  
(Admittedly, this ``standard'' method has not yet
been used much for interpreting data, but this may change soon.) 
After my redefinitions we have only four parameters, with much more cleanly 
separated effects:
\begin{eqnarray*}
{\rm standard~ parameters} &\longrightarrow & 
\rho_0,~\rho_0^\prime,~\rho_0^{\prime\prime},~\rho_0^{\prime\prime\prime},~
N_0,~{\rm and~ messy~ cutoffs} \nonumber \\
{\rm new~ parameters} &\longrightarrow & 
\bar{\rho}_g,~b_1,~\tilde{b}_2,~{\rm and}~ N~.
\end{eqnarray*} 
\cite{1998MNRAS.301..797H} provided the core calculations for this paper.  
The innovation here is the suggestion that their effective linear bias and 
shot-noise-like term should not be treated as output predictions of the 
model but 
instead should be absorbed into renormalized linear bias and shot-noise 
parameters.
Significantly, these new free parameters of the model contain most of the 
high-$k$ (small, non-linear scale) sensitivity in the bias calculation.  
Rather than having a large-scale effective bias (ratio of tracer power to mass 
power) which is a function of all three bias parameters in the model, and the 
smoothing scale, I have only 
a single parameter for the large-scale bias.  
Rather than shot-noise that is a 
function of the
original stochasticity, the 2nd order bias, and the smoothing scale, I have a 
single 
shot-noise parameter that describes all asymptotically large-scale deviation 
from the linear bias model.  Quasi-linear deviation from the linear bias plus 
shot-noise model is described by a single parameter, with no need for 
smoothing.

One might ask if it is desirable to bury time and cosmology dependence, in the 
form
of the values of $\sigma^2$ and $\int dq~ q^2 P(q)^2$, in $b_1$ and $N$.
Do we not lose information this way? The answer is: not really, because 
$c_1$, $c_2$, $c_3$, and $N_0$ are also mixed into $b_1$ and $N$, and, even
for fixed smoothing scale, generally none of these parameters will be 
independent
of the cosmological model or of time.  Furthermore, $\sigma^2$ and 
$\int dq~ q^2 P(q)^2$ are 
relatively sensitive to high-$k$ power which is poorly described by 
perturbation theory.

Note that simulations or other microscopic models of galaxy formation can still
be used to calibrate the renormalized large-scale structure model.  The way 
this works exemplifies the usefulness of the revised model.  The most 
straightforward
method for calibrating the unrenormalized model is to pick some smoothing scale
and measure directly the parameters of the original Taylor series, 
Eq. (\ref{eqtaylor}).  If this is done very literally
(e.g., by making a scatter plot of galaxy density vs. smoothed mass density
and reading off derivatives at $\delta=0$), one is not even
guaranteed that the calibrated model will reproduce the mean density of 
galaxies in the 
simulations used to calibrate it (without the trivial renormalization of the
first term in the Taylor series,
discussed above), let alone the large-scale bias, the 
shot-noise, or really anything else about large-scale structure.  (I emphasize
here that I am not saying this method will surely fail, only that it is not 
guaranteed to work.)
To calibrate
the renormalized model one computes the mean density and power spectrum from
the simulation, and
if desired the bispectrum, and then fits for the parameters of the model.
This guarantees that the large-scale structure in the simulation will be 
reproduced by the model as well as possible.  Of course, one can be less 
literal and calibrate the unrenormalized model by fitting the simulated 
statistics,
but this still includes the extra steps of choosing a smoothing scale and 
computing the simple observables (e.g., mean density, large scale bias) as a 
function
of several parameters.  Of course, as I discussed above, there is an additional 
problem with the 
literal use Eq. (\ref{eqtaylor}) in that any smoothing aggressive enough to 
render the density fluctuations small will also corrupt the power on  
interesting, perturbative scales.

Within the perturbative approach there is a simulation-independent means of 
estimating the
validity of the results.  One can go through the full process of extracting the
cosmological parameters of most fundamental interest from the data, 
marginalizing over the bias parameters, using 
the perturbative model calculated to two different orders, and compare the 
results.  As presented here, the results from linear theory could be compared 
to the model with
the first non-linear corrections, although computing to another order might be 
useful in the future (of course, higher order is necessary if one wants to use
the bispectrum to two different non-zero orders).  
The level of agreement would surely depend on the 
maximum $k$ used in the fit.  This maximum $k$ could be chosen small enough to 
ensure agreement, i.e., consistency of the perturbative model.    

There are several obvious next steps for this line of work:
Extension to redshift space \cite{2003ApJ...585...34M,2004PhRvD..70h3007S} 
should be 
straightforward, following the 
calculations of \cite{1998MNRAS.301..797H}.
A more general time and space-local model could be implemented, e.g.,
once one goes beyond linear theory the Taylor series for $\rho_g$ should 
probably include velocity divergence terms as well as density terms.   
As discussed by \cite{2006astro.ph..6028M}, it may be possible to make 
non-trivial predictions about the time evolution of bias by starting the 
perturbative calculation from a model for the local formation (and merging,
etc.) rate of the 
tracer \citep{1998ApJ...500L..79T,2005A&A...430..827S}, 
rather than the instantaneous density of the tracer (this will 
probably require a renormalization group approach as in 
\cite{2006astro.ph..6028M}, rather than simple parameter redefinitions as in
this paper).
Extension to multiple luminosities or types of galaxies (or more divers 
tracers \cite{2000ApJ...537...37T,
2002ApJ...580...42M,2006MNRAS.369...68S}) can be done by writing
a Taylor series like Eq. (\ref{eqtaylor}) for each.  This will lead to extra 
parameters, but cross-correlation will provide extra observables to constrain
them. 

Finally, I
observe that, beyond just removing badly behaved perturbative terms at the 
present order of 
calculation, by treating the linear bias and shot-noise as free parameters
we are in effect already accounting for similar terms that appear at any 
order in perturbation theory.  
This is probably the most compelling reason 
to believe that the approach of this paper can be expected work 
better than the standard approach as presented (the problem in the standard 
approach of 
smoothing corrupting the power on interesting scales is also significant).  
It may be possible to show that all higher
order corrections to the tracer power as $k\rightarrow 0$ can be included in 
these
two terms, i.e., that the renormalized linear bias and shot-noise give a
fundamentally complete descriptions of very large-scale clustering
\cite{1998ApJ...504..607S}. 


\acknowledgements

I thank Roman Scoccimarro for helpful comments on the manuscript, Chris 
Hirata for a helpful conversation, and Eiichiro Komatsu for pointing out 
an error in the original versions of Eq. (\ref{eqbarepower}) and 
(\ref{eqfinalpower}).
I thank Juan Garc\'ia-Bellido, Enrique Gazta\~naga, Julien Lesgourgues, 
David Wands, and Maria Beltran for organizing the Benasque Workshop on 
Modern Cosmology where much of this work was completed. 
Some computations were performed on CITA's
McKenzie cluster which was funded by the Canada Foundation for Innovation and
the Ontario Innovation Trust \citep{2003astro.ph..5109D}.

\bibliography{cosmo,cosmo_preprints}

\end{document}